\documentclass[letterpaper]{article} 
\usepackage[submission]{aaai25}  
\usepackage{times}  
\usepackage{helvet}  
\usepackage{courier}  
\usepackage[hyphens]{url}  
\usepackage{graphicx} 
\usepackage{amssymb}
\usepackage{natbib}  
\usepackage{multirow}

\usepackage{caption} 
\usepackage{algorithm}
\usepackage{algorithmic}
\usepackage{amsmath}
\usepackage{color}

\usepackage{xcolor}

\usepackage{hyperref}
\usepackage{newfloat}
\usepackage{listings}
\usepackage{multicol}
\DeclareCaptionStyle{ruled}{labelfont=normalfont,labelsep=colon,strut=off} 
\floatstyle{ruled}
\newfloat{listing}{tb}{lst}{}
\floatname{listing}{Listing}
\title{\includegraphics[width=1.6cm,height=0.4cm]{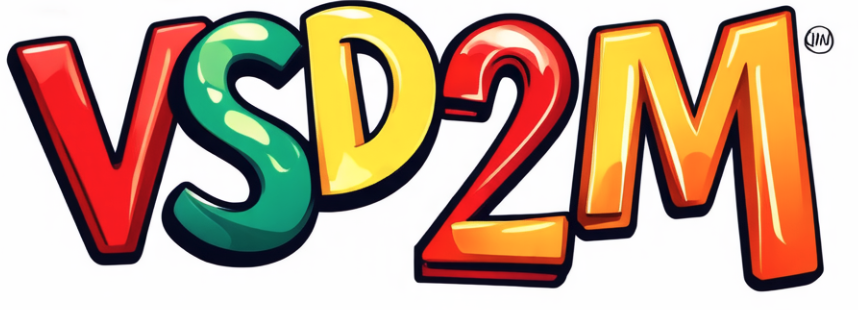} : A Large-scale Vision-language Sticker Dataset for \\ Multi-frame Animated Sticker Generation}

\author{
    Zhiqiang Yuan\textsuperscript{\rm +}, Jiapei Zhang\textsuperscript{\rm +}, Ying Deng, Yeshuang Zhu, Jie Zhou, Jinchao Zhang\thanks{corresponding author, \textsuperscript{+} equal contribution.} \\
    Wechat AI, Tencent
}
\affiliations{
    \textsuperscript{\rm 1}Association for the Advancement of Artificial Intelligence\\

    1101 Pennsylvania Ave, NW Suite 300\\
    Washington, DC 20004 USA\\
    proceedings-questions@aaai.org
}

\usepackage{bibentry}

\begin{document}

\maketitle

\begin{abstract}
    As a common form of communication in social media,
    stickers win users’ love in the internet scenarios, for their ability to convey emotions in a vivid, cute, and interesting way.
    People prefer to get an appropriate sticker through retrieval rather than creation for the reason that creating a sticker is time-consuming and relies on rule-based creative tools with limited capabilities.
    Nowadays, advanced text-to-video algorithms have spawned numerous general video generation systems that allow users to customize high-quality, photo-realistic videos by only providing simple text prompts. 
    However, creating customized animated stickers, 
    which have lower frame rates and more abstract semantics than videos, is greatly hindered by difficulties in data acquisition and incomplete benchmarks.
    To facilitate the exploration of researchers in animated sticker generation (ASG) field, 
    we firstly construct the currently largest vision-language sticker dataset named ``VSD2M'' at a two-million scale that contains static and animated stickers.
    Secondly, to improve the performance of traditional video generation methods on ASG tasks with discrete characteristics, we propose a \textbf{S}patial \textbf{T}emporal \textbf{I}nteraction (STI) layer that utilizes semantic interaction and detail preservation to address the issue of insufficient information utilization.
    Moreover, we train baselines with several video generation methods (\textit{e.g.,} transformer-based, diffusion-based methods) on VSD2M and conduct a detailed analysis to establish systemic supervision on ASG task. 
    To the best of our knowledge, this is the most comprehensive large-scale benchmark for multi-frame animated sticker generation, and we hope this work can provide valuable inspiration for other scholars in intelligent creation.
   Our dataset and code will be released at \textcolor{blue}{\href{https://xiaoyuan1996.github.io/files/VSD2M/index.html}{link}}.
\end{abstract}

%

\section{Introduction}
Stickers, as a commonly used medium on social platforms, play the role of ``a picture is worth a thousand words'' in dialogue and communication \cite{lei2016rating}\cite{ huang2017overlapping}.
Unlike generic scenes, stickers usually contain both real and cartoon forms, with rich emotions and optical characters, which bring great challenge for models in terms of comprehension and modeling.



As a subset of stickers, multi-frame animated stickers (GIFs) are more popular because of their interesting actions and imaginative divergence.
Compared to video, GIFs have a lower frame rate, which results in low pixel smoothness between frames, making the temporal information more discrete and abstract.
Due to the difficulty in data acquisition and imperfect benchmarks, relevant research \cite{zhao2023sticker820k} on GIFs is relatively rare to the best of our knowledge.
Existing sticker-related works \cite{liu2022ser30k, soujanya2018multimodal, zhao2023sticker820k} currently follow two drawbacks:
(1) The relevant open-source data set only contains static stickers with a small data size.
(2) The previous works all focus on the sticker understanding, but the generation of stickers, especially animated stickers, is not involved.
The above problems have greatly restricted the development of customized sticker generation, which has an important application in human-computer interaction.

\begin{figure*}[!t]
\centering
\includegraphics [width=6.7in]{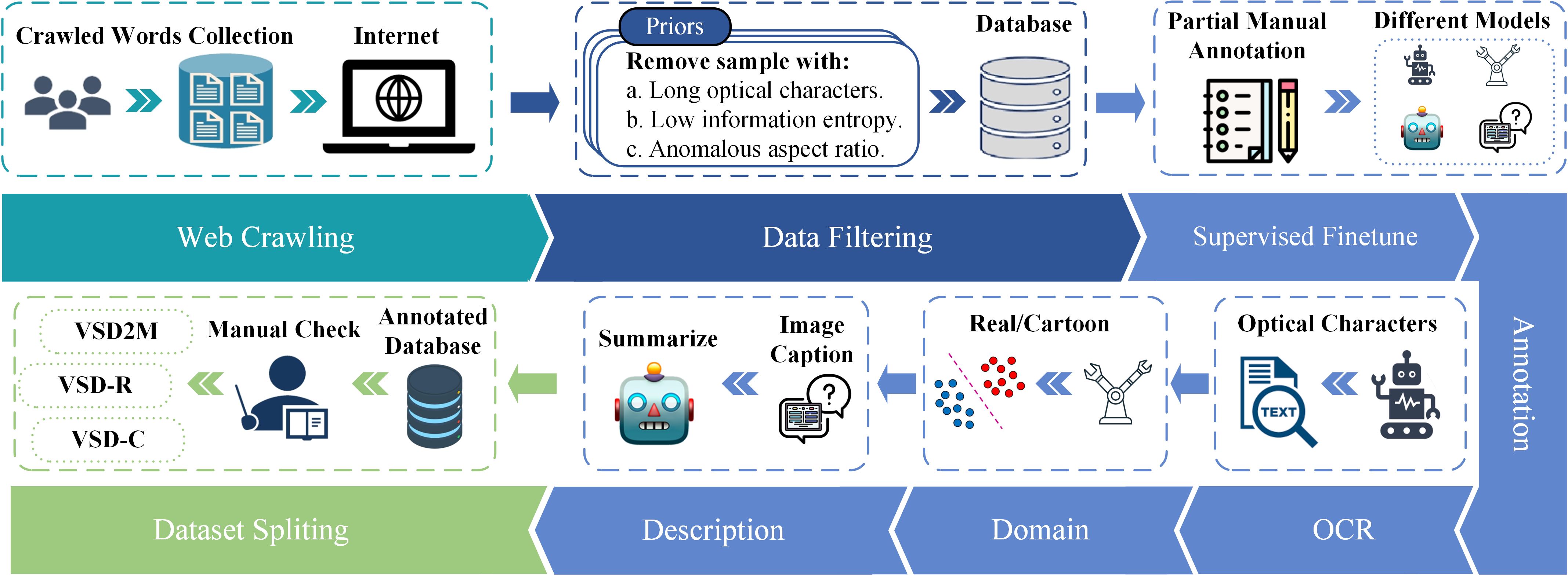}
\captionsetup{font={small},skip=4pt, justification=raggedright}
\caption{
Overview of data collection and processing, which can be divided into four stages: web crawling, data filtering, annotation and dataset splitting.
During the data annotation process, we use manually labeled data to fine-tune different models to obtain high-quality semi-automatic annotation results.
}
\label{pipeline}
\end{figure*}

Apart from the lack of data, there is also a scarcity of methods when modeling ASG tasks, which forces us to leverage other generation methods to solve this problem.
As a generation task similar to ASG, video generation has made remarkable progress in recent years, however, utilizing it to achieve high-quality animated stickers is nontrivial:
(1) When the data is relatively discrete, it is difficult for the current video generation methods to handle entities with significant changes across different frames, due to their small spatial receptive fields in temporal modeling.
(2) Video generation methods typically aim at high frame rate data, allowing the model to rely on recent frames for better next-frame predictions, while when changes increase, more frames are needed for next-frame prediction.
The above two issues are caused by insufficient utilization of spatial and temporal information, which jointly limits the application of video generation methods in modeling discrete data such as animated stickers.

Driven by the above issues, this paper
shows a complete benchmark to promote the development of animated sticker generation task.
\textbf{Firstly}, 
we construct a large-scale vision-language sticker datasets with two-million samples, which is about three times the size of the current largest sticker dataset Sticker820k \cite{zhao2023sticker820k}.
In addition to caption and static stickers, VSD2M also contains GIFs that are not included in other works, and the annotated caption provides a detailed description to the relevant action in sticker GIFs.
To promote academic research by relevant scholars, we make the data open to access in order to lay the groundwork for the ASG filed.
\textbf{Secondly},
we propose a spatial temporal interaction layer for discrete sticker generation, which divides traditional temporal modeling into semantic interaction and detail preservation so as to alleviate the insufficient utilization of information.
The STI layer constructs semantic interactions across spatial and temporal dimensions, while also preserving image details, thereby alleviating potential issues of traditional video generation methods in modeling discrete data.
\textbf{Thirdly},
we explore the performance of different video generation methods (\textit{e.g.,} transformer-based, diffusion-based methods) on animated sticker generation tasks.
We conduct detailed experiments, provide exhaustive 
results, and release solid benchmarks to facilitate ASG task.
We aim to offer inspiration and guidance to relevant scholars, as well as to provide data support for this field.

The main contributions of our work are as follows:
\begin{itemize}

    \item We provide a complete benchmark for animated sticker generation.
    Within our knowledge, this represents a well-established effort towards the intelligent creation of animated stickers.

    \item We construct a million-level vision-language sticker dataset containing both static and animated stickers, which is the largest and most comprehensive multi-modal sticker dataset to date.


    \item A spatial-temporal interaction layer is proposed for discrete sticker generation, which alleviates the insufficient information of generation methods in ASG tasks through semantic interaction and detail preservation.

\end{itemize}

\section{Related Works}
\label{related_workds}

\subsection{Vision-Language Datasets}

\begin{figure*}[!htbp]
\centering
\includegraphics [width=6.7in]{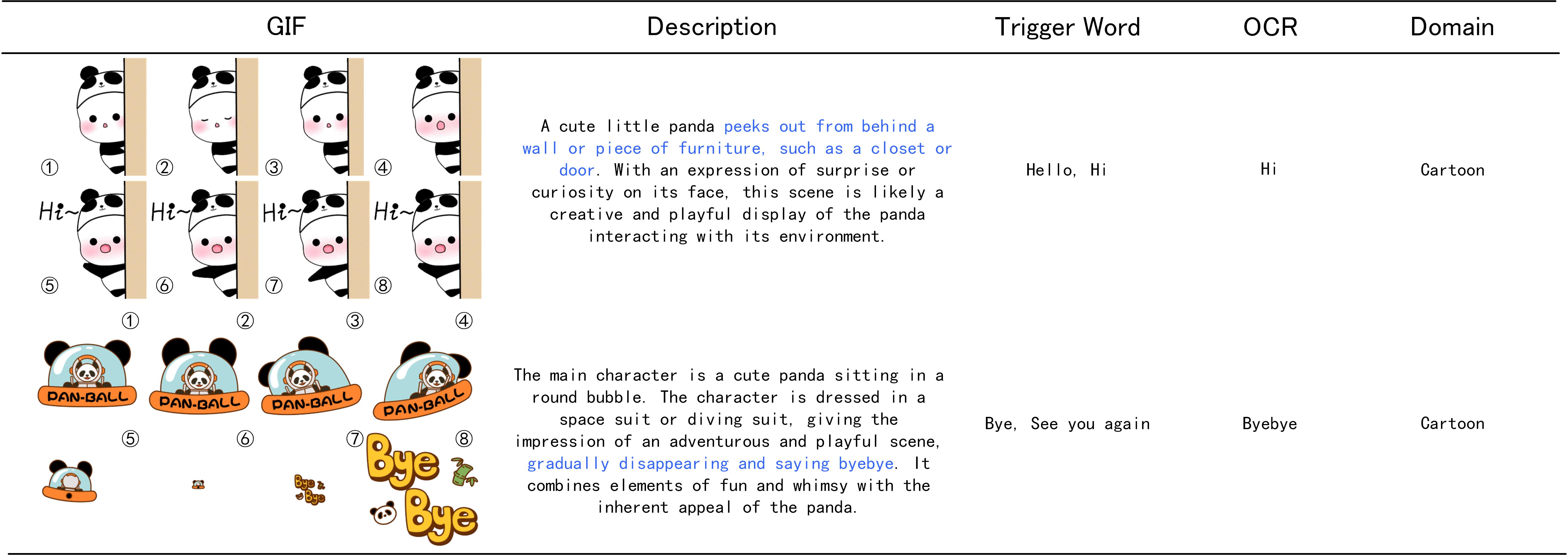}
\captionsetup{font={small},skip=4pt, justification=raggedright}
\caption{Two samples of VSD2M, in which GIFs is framed for visualization.
The \textcolor[RGB]{47,85,150}{blue} description shows part of the action in the GIFs.}
\label{data_show}
\end{figure*}

\textbf{Natural Vision-language Datasets.}
Recently, visual-language tasks including cross-modal retrieval \cite{zhang2022contrastive, li2023blip}, contents generation \cite{zhu2023minigpt, luo2023videofusion, blattmann2023align}, $etc.$ have made considerable progress, which is closely related to the establishment of large-scale multi-modal datasets.
Most of these datasets mainly contain text-image pairs in natural scenes, such as Flickr30k \cite{plummer2015flickr30k}, MSCOCO \cite{lin2014microsoft}, Visual Genome \cite{krishna2017visual}, LAION-400M \cite{schuhmann2021laion}, LAION-5B \cite{schuhmann2022laion}.
In addition, there are also many datasets containing video modality \cite{bain2021frozen, brooks2022generating, siarohin2019first, xiong2018learning}, which are often used for video generation \cite{luo2023videofusion, blattmann2023align} and interpretation \cite{maaz2023video, zhang2023video}.
The above datasets greatly facilitate the pre-training of multi-modal models in natural scenes, allowing them to achieve well performance in downstream tasks through fine-tuning.

\noindent
\textbf{Sticker Vision-language Datasets.}
Different from various Visual-Language Datasets (VLD) in natural scenes, VLD is relatively rare in the sticker field, which acts a common interactive tool for humans \cite{poria2018meld, gao2020learning, fei2021towards}.
Considering the lack of sticker data, Liu $et\ al.$ \cite{liu2022ser30k} constructed a large-scale dataset for sticker emotion recognition, named SER30K, which includes 30K static stickers with the emotion categories from 1887 topics. 
Further, Zhao $et\ al.$ \cite{zhao2023sticker820k} collected an sticker-text dataset Sticker820K, containing 820K samples to perform sticker-related retrieval and caption generation tasks.
However, the above works were all performed on static stickers rather than animated stickers, which lacking dynamic action in streaming.
For this reason, we have established a million-level multi-frame animated sticker dataset to facilitate the development of multi-modal sticker-related visual-language generation.

\subsection{Video Generation}

Video generation is an urgent task in advanced artificial intelligence.
Most of the earlier video generation methods \cite{saito2017temporal, skorokhodov2022stylegan} are based on adversarial networks \cite{creswell2018generative}, which achieve well temporal consistency by allowing the generator to directly learn the joint distribution of video frames.
With the development of self-attention mechanism \cite{vaswani2017attention}, researchers manage to utilize transformers to model video generation tasks and achieve acceptable computing efficiency and scalability \cite{yan2022patch, hong2022cogvideo}.
In recent years, due to the key advances of Diffusion Probabilistic Models (DPMs) in generation tasks, researchers have attempted to use DPMs for video generation \cite{ yang2022diffusion, ni2023conditional}.
Most studies optimize this task from DPMs' architecture \cite{luo2023videofusion, ge2023preserve, blattmann2023align} and controllability \cite{wu2023tune, wang2023videocomposer, he2023animate}, thereby achieving well generation performance in natural scenarios.
The above models for general fields have achieved great success, however as shown in Table \ref{sticker_review}, current researchers in the sticker field have not conducted relevant explorations into ASG task.
In this paper, we leverage
the transformer-based and diffusion-based video generation approaches to 
establish a set of solid baselines so as to provide exhaustive benchmarks for ASG tasks.

\begin{table}[!t]
 \small
\resizebox{\linewidth}{!}{
\begin{tabular}{cccc}
\hline
Works    & Dataset     & Task                          \\ \hline
\cite{poria2018meld}   & MELD        & Sticker Emotion Recognition   \\
\cite{gao2020learning}      & SRS         & Sticker Response Selector     \\
 \cite{fei2021towards}     & MOD         & Sticker Response Selector     \\
 \cite{sheynin2022knn}  &  KnnSticker           & Static Sticker Generation     \\
 \cite{liu2022ser30k}    & SER30K      & Sticker Emotion Recognition   \\
 \cite{zhao2023sticker820k}     & Sticker820K & Sticker Retrieval and Caption \\ \hline
\end{tabular}
}
\caption{
The datasets and the corresponding tasks proposed by recent sticker-related works.
Although there are some tasks involving stickers, researchers have not further explored the task of generating animated stickers.
}
\label{sticker_review}
\end{table}

\section{VSD2M DATASET}
\label{mvsd}

\subsection{Data Construction}

In this subsection, we construct a million-level vision-language sticker dataset to facilitate the ASG task.

Fig.\ref{pipeline} show the overview of data collection and processing.
We initially obtained 2.5M data from Internet
by retrieving 600K crawled words, and then the samples that meet the following conditions were removed based on observed priors: 
\textit{
a. The sample with long optical characters.
b. The sample with low information entropy.
c. The sample with anomalous aspect ratio.
}
After this, the candidate data which contains 2.1M static and animated stickers are obtained.

\begin{figure}[!h]
\centering
\includegraphics [width=3.2in]{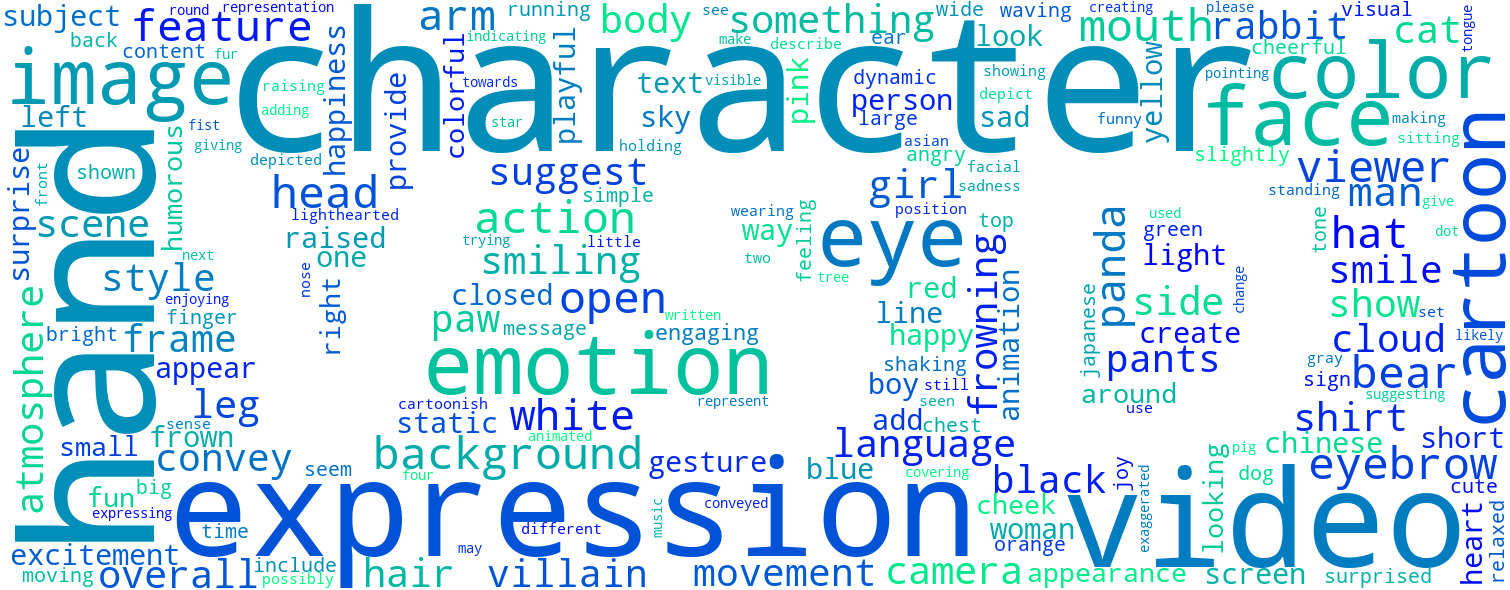}
\captionsetup{font={small},skip=4pt, justification=raggedright}
\caption{Word cloud distribution of the description in VSD2M,
which contains information that reflects the motion in GIFs, such as \textit{movement}, \textit{open}, $etc$.
}
\label{cloud_freq}
\end{figure}

\begin{figure*}[!t]
\centering
\includegraphics [width=6.7in]{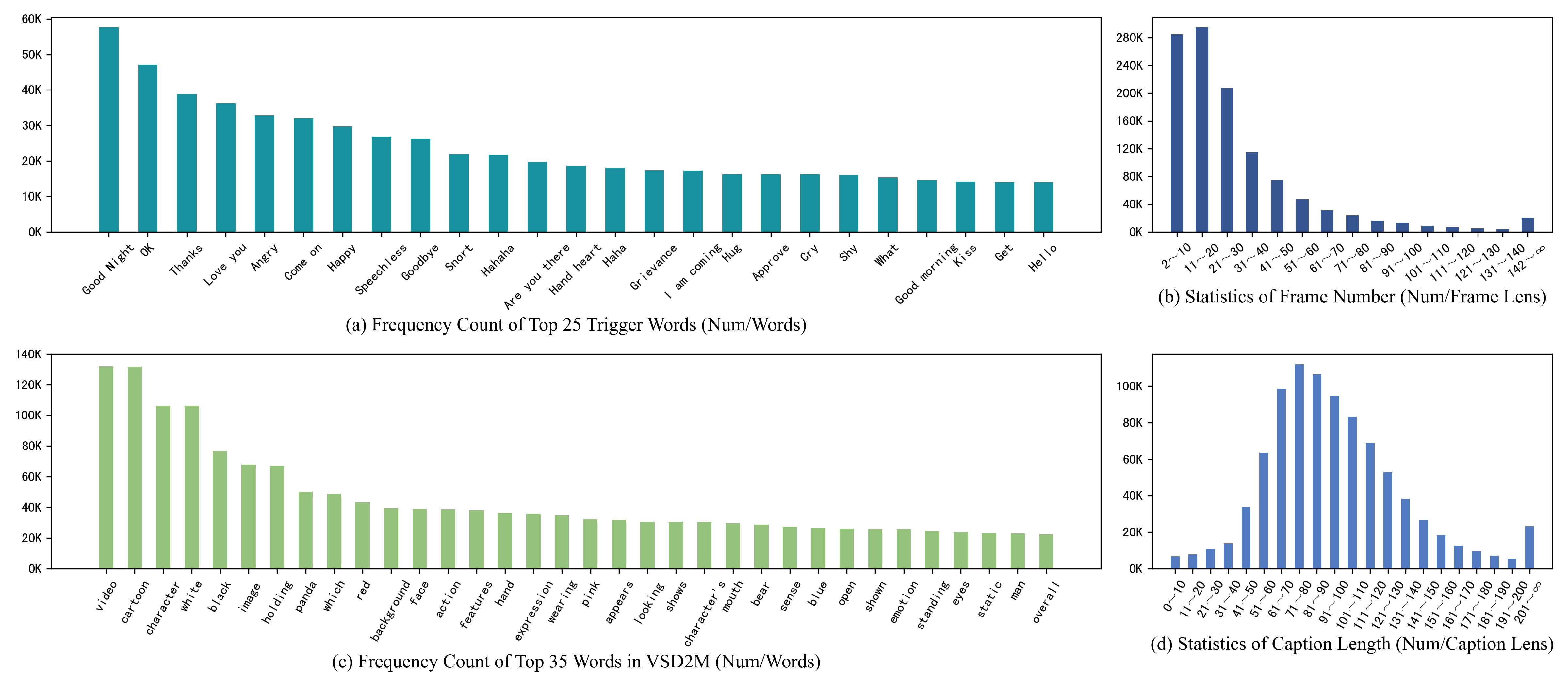}
\captionsetup{font={small},skip=4pt, justification=raggedright}
\caption{
Visual analysis of VSD2M.
(a) Frequency count of top 25 trigger words. 
(b) Statistics of frame number, note that we only count multi-frame animated stickers.
(c) Frequency count of top 35 words in descriptions.
(d) Statistics of caption length.
}
\label{data_freq}
\end{figure*}

\begin{table*}[!t]
\centering
\resizebox{\linewidth}{!}{
\begin{tabular}{c|ccccccc}
\hline
Dataset     & Size & Modality &     language      & Trigger Word & OCR & Real/Cartoon Flag & Is Open \\ \hline
TGIF \cite{li2016tgif}  & 100K & GIFs, description & En & & & &  \checkmark \\
MOD \cite{fei2021towards}      & 45K    & static sticker &                      &        &      &     & \checkmark
\\
SRS \cite{gao2020learning}      & 320K    & static sticker &      &       &      &   & \checkmark             \\
SER30K \cite{liu2022ser30k}      & 30K    & static sticker & En                     & \checkmark       &      &        \checkmark     & \checkmark    \\
Sticker820K \cite{zhao2023sticker820k} & 820K   & static sticker, description & En     & \checkmark        & \checkmark   &        &       \\
VSD2M        & \textbf{2.09M}    & \textbf{static sticker, GIFs, description} & En, \textbf{Cn}  & \checkmark       & \checkmark     & \textbf{\checkmark}  & \checkmark              \\ \hline
\end{tabular}
}
\caption{
Static information comparison of different vision-language sticker datasets.
}
\label{datasets_comp}
\end{table*}

To reduce the semantic gap between modalities, several different strategies are utilized to enrich the dataset labels.
We manually labeled portions of the data for different annotation tasks, allowing us to fine-tune the following models.
\textbf{a. OCR.}
PaddleOCR \cite{du2020pp} finetuned on the labeled sticker dataset is used to extract optical characters from visual samples.
\textbf{b. Domain.}
We train a classification model based on EfficientNet \cite{koonce2021efficientnet} to provide discrimination between real and cartoon scenes.
\textbf{c. Description.}
For GIFs, we use 330K pieces of manually annotated data to perform supervised fine-tuning on VideoLlama \cite{zhang2023video} to improve the accuracy of automatic annotation, so that the trained model can be directly used for visual descriptions generation.
Different from Sticker820K \cite{zhao2023sticker820k}, the description in VSD2M contains the action information of the GIFs, which has an important impact on sticker generation control \cite{yang2020captionnet}.
While for static stickers, we use LLaVa \cite{liu2023visual} to extract description.
After obtaining descriptions of static and animated stickers, we use a large language model \cite{yang2023baichuan} to produce cleaner captions by summarizing the results, where the prompt leveraged is:
``\texttt{Summarize the following text from both static and dynamic features: \textbackslash n \textcolor{orange}{[Description]}}''.
In addition to the English version, we also provide a Chinese version to facilitate the needs of researchers.
\textbf{d. Trigger word.}
Crawled words related to the visual sample are aggregated and regarded as trigger words.
The comparison of static information between the VSD2M and other sticker-related datasets is shown in Table \ref{datasets_comp}.
VSD2M incorporates GIFs modality compared to other sticker datasets and is greatly ahead in terms of data scale.

\begin{table}[!t]
\resizebox{\linewidth}{!}{
\begin{tabular}{ccc}
\hline
Multi-frame Ratio    &  Optical Character Ratio   &   Ave Frames Numbers       \\ \hline
51.9\%           & 55.2\%            & 16.91                       \\ \hline
Cartoon Ratio & Ave Trigger Words  & Ave Description Length \\ \hline
45.6\%              & 36.10           & 104.01                     \\ \hline
\end{tabular}}
 \caption{Numerical properties of the VSD2M dataset.}
\label{numerical_prop}
\end{table}

In order to facilitate the use of researchers, we manually select from multiple pieces of samples corresponding to each top 500 trigger words in real and cartoon domain as test sets, named as VSD-R and VSD-C respectively.
Subsequent experiments on these two test sets are utilized to evaluate the model performance on different domains.

\subsection{Data Statistics}

Two examples of the VSD2M dataset are shown in Fig.\ref{data_show}, in which each sample contains a animated or static sticker, a description, a set of trigger words, OCR results, and domain identification (Cartoon/Real).
In the annotated description, in addition to static content such as ``black veins'', the actions such as ``swaying in the wind'' is also well represented.
The annotated caption has more image details and can cover rich visual semantics.
The Fig.\ref{cloud_freq} shows the word cloud distribution of the annotated description, which shows that the most used words include
\textit{video}, \textit{expression}, \textit{cartoon}.
Table \ref{numerical_prop} shows some numerical properties of VSD2M, including the proportion of cartoon samples, the average number of frames, $etc$, in which real and cartoon domain both account for nearly half.
To further show the details of the dataset, we also count the distribution of static attributes such as trigger words, caption length, $etc.$ in Fig.\ref{data_freq}.
VSD2M aims to provide rich data to the scholars in intelligent creation, thus facilitating the related research in sticker generation. 

\section{Algorithmic Analysis}


This section briefly introduces the current video generation methods, then proposes the STI layer to handle discrete data generation, and finally gives a solid baseline for ASG task.

\label{Algorithmic}

\subsection{Methodology in Video Generation}


We select four state-of-the-art models from transformer-based and diffusion-based video generation methods, where the difference between the two architectures is shown in Appendix \textcolor{red}{A.1}.
Assuming a paired dataset with samples $(\boldsymbol{x},\boldsymbol{y}) \sim p_{\text{train}}(\boldsymbol{x},\boldsymbol{y})$, where $\boldsymbol{x}$ denotes the video clip with $N$ frames and $\boldsymbol{x}=\{x^i | i=1,2,...,N\}$, $\boldsymbol{y}$ is the corresponding caption.

\noindent
\textbf{VideoGPT \cite{yan2021videogpt}} is a vision transformer framework that leverages masked token prediction and prior learning.
The framework consists of two stages.
Firstly, VideoGPT leverages a encoder to tokenize the video as $f_{\mathcal{T}}: \boldsymbol{x} \xrightarrow{} z \in \mathbb{Z}$ by a 3D-VQ network, where $\mathbb{Z}$ is a codebook.
After video modeling, the decoder $f_{\mathcal{T}}^{-1}$ maps the latent tokens back to video pixels.
In the second stage, VideoGPT learns a prior over the latent codes from the first stage.
This method embeds conditions as corrupted visual tokens into multivariate mask $\textbf{m} \sim \mathcal{P}_{\mathcal{U}} $, thus to obtain and recover the corrupted input $\overline{z} = \textbf{m}(z)$\cite{huang2024wavedm}.
The training objective is
\begin{equation}
    \mathcal{L} = \mathbb{E}( 
    -log\, p_{\theta} (z | [\boldsymbol{y}, \overline{z}])
    )
\end{equation}
where $\theta$ is the learnable parameters and $\mathbb{E}$ is the mathematical expectation.

\noindent
\textbf{Make-A-Video} is based on the diffusion probability models.
DPMs aims to sample from the distribution $p(x^i)$ by denoising and converting samples from a Gaussian distribution into target distribution iteratively \cite{sohl2015deep, song2020score}, which can be divided into diffusion process and reverse process.
In step $t$ during diffusion process, a noised image $x^i_t$ is generated by $x^i_t = \sqrt{\alpha_t} x^i + \sqrt{(1-\alpha_t)} \epsilon$, $\epsilon \sim \mathcal{N}(0, \boldsymbol{I}_d)$, where $\epsilon$ is a Gaussian noise and $\alpha_t$ controls the noise ratio at step $t$.
In the reverse process, a learnable network $\mathcal{G}_{\theta_1}(x^i_t, t)$ aims to predict the noise and recover the clean image from the noised input $x^i_t$.
After training, the model starts from a pure Gaussian noise $x_T \sim \mathcal{N}(0, \boldsymbol{I}_d)$ and samples a clean image by iteratively running for $T$ steps.
Make-A-Video regards the caption $\boldsymbol{y}$ as guidance information to generate the paired video clip $\boldsymbol{x}$.
In this case, the model samples from the conditional distribution of $p(\boldsymbol{x}|\boldsymbol{y})$, and the learnable network $\mathcal{G}_{\theta_2}
([\boldsymbol{x}_t, \boldsymbol{y}], t)$ parameterized by $\theta_{2}$ utilize the $\boldsymbol{x}_t$ and $\boldsymbol{y}$ as input.
Make-A-Video adds 3D convolutional layers to the pre-trained image generation model, so that the model can obtain temporal modeling capabilities and produce consistent visual semantics between frames.
During training, the loss function can be modeled as:
\begin{equation}
    \mathcal{L}_t=
     \mathbb{E}( \mathcal{G}_{\theta_2}( \boldsymbol{y}, \sqrt{\alpha_t} \boldsymbol{x}+ \sqrt{1-\alpha_t} \epsilon_t, \alpha_t ) - \epsilon_t )
\label{eqa:diffusion-loss}
\end{equation}
where $\mathcal{L}_t$ is the loss function and $\epsilon_t$ is the noise at step $t$.


\begin{figure*}[!t]
\centering
\includegraphics [width=6.9in]{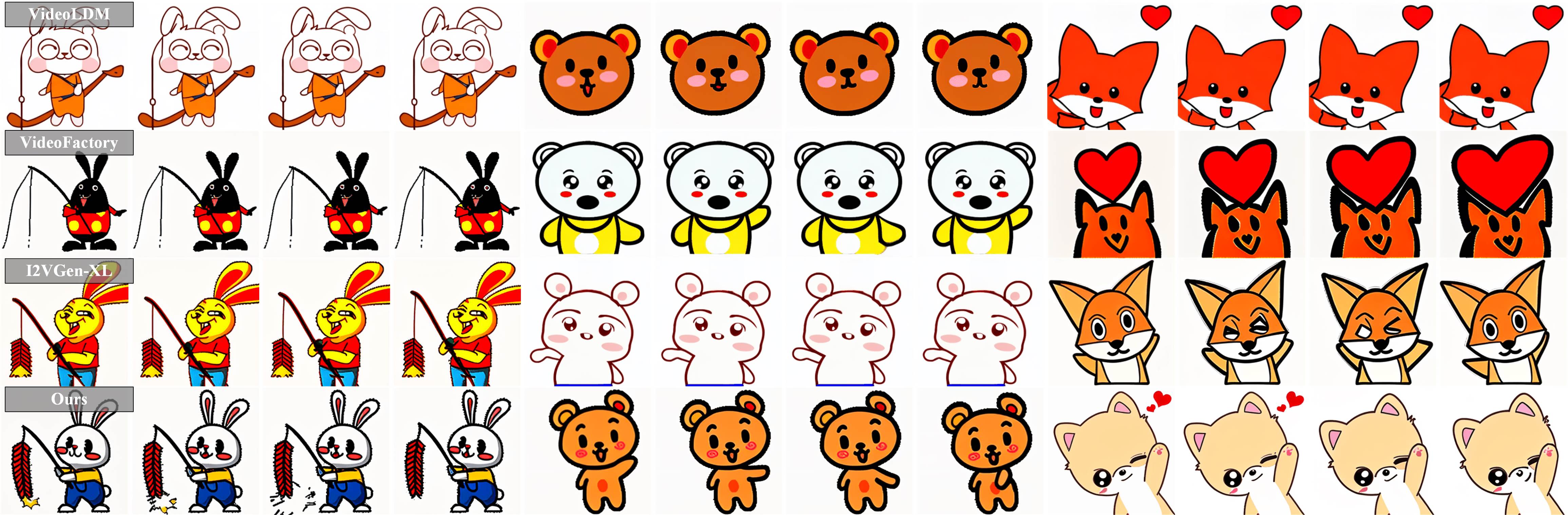}
\captionsetup{font={small},skip=4pt, justification=raggedright}
\caption{
Visual comparison for animated sticker generation between VideoLDM, VideoFactory, I2VGen-XL and ours.
The text prompts are as follows,
left: ``A cute rabbit setting off firecrackers'',
middle: ``A little bear waving his hands up and down'',
right: ``A cartoon little fox waving with a heart above his head''.
More results can be seen in \textcolor{blue}{\href{https://xiaoyuan1996.github.io/files/VSD2M/index.html}{[link]}}.
}
\label{comp_cartoon}
\end{figure*}

\noindent
\textbf{VideoLDM \cite{blattmann2023align}} is also the diffusion-based method, which converts the generation process to latent space so as to accelerate the training process through encoder and decoder.
This model adds the adaptive spatial and temporal fusion based on the Make-A-Video \cite{singer2022make} to enable the interaction of different modalities.
More specifically, it sets trainable parameters to balance the contribution of 2D and 3D information during modeling, thus achieving well results in video generation tasks such as driving video generation.

\noindent
\textbf{VideoFactory \cite{wang2023videofactory}} argues that Previous approaches \cite{blattmann2023align} utilize temporal 1D convolution/attention modules to integrate temporal information, which overlook the importance of jointly modeling space and time.
To enhance the temporal consistency and alignment between texts and videos, VideoFactory introduces a swapped cross-attention mechanism in 3D windows to enable the better interaction between spatial and temporal perceptions.

\noindent
\textbf{I2VGen-XL \cite{zhang2023i2vgen}}
propose a cascaded approach that enhances model performance by decoupling these two factors and ensures the alignment of the input data by utilizing static images as a form of crucial guidance.
In this paper, we discard its image guidance and only use text guidance to make a fair comparison with other models.

\subsection{STI Layer for Discrete Sticker Generation}

Although current video generation methods have achieved remarkable results, modeling discrete data is nontrivial.
Two key problems are as follows:
\textbf{a.} 
When the data is relatively sparse, the same entity exhibits considerable variation across different frames.
However, most current works employ convolution kernels with size $1\times1$ in the spatial layers when modeling temporal connections, and the small receptive field makes it hard to obtain the changing relationships of entities in adjacent frames.
\textbf{b.} 
Current video generation methods are typically designed for generating high frame rate data. In this case, the model only needs to consider the most recent frames to make accurate predictions for the next frame.
But in conditions where data is sparse, the model needs to consider more frames.
The two problems above are caused by insufficient utilization of information at spatial and temporal level respectively, which jointly restrict the application of video generation methods in discrete data modeling, such as multi-frame stickers.

To deal with the above issues, we propose a \textbf{S}patial \textbf{T}emporal \textbf{I}nteraction (STI) layer for discrete sticker generation, which further divides temporal modeling into semantic interaction and detail preservation so as to alleviate the insufficient utilization of information.
\textbf{a. Semantic interaction.}
To selectively focus on regions in other frames when predicting the current frame, we attempt to interact the region features in all frames during temporal modeling.
Specifically, for the feature $h \in \mathbb{R}^{F\times H\times W\times d }$ that needs to be sent to the temporal layer, where ${F, H, W, d}$ represent frame, height, width, and dimension respectively, we first downsample it by $\gamma$ times to reduce the interaction complexity and name it as $h_{\gamma} \in \mathbb{R}^{F\times \frac{H}{\gamma}\times \frac{W}{\gamma}\times d }$.
Next, we unfold its temporal and spatial dimensions $ \mathbb{R}^{F\times \frac{H}{\gamma}\times \frac{W}{\gamma}\times d }\rightarrow \mathbb{R}^{F \frac{H}{\gamma} \frac{W}{\gamma}\times d }$, and utilize self-attention to achieve interaction between different frames.
It should be noted that in GIFs scenes with long frames, combined with appropriate $\gamma$ values, the complexity of self-attention operations will not increase significantly.
After that, we use upsampling with the same magnification of $\gamma$ and tensor transformation  to restore the transformed $h_{\gamma}$ back to the original size.
\textbf{b. Detail preservation.}
Although semantic interaction can take into account the global features of GIFs for next-frame prediction, it may lead to a loss of image details.
To preserve details, we use convolutions with kernel size $k\times1\times1$ to reduce the attenuation of fine-grained features, and $k$ is the kernel size of frame.
We extract features with different patterns and then set learnable weights to balance the proportion of these two branches. This allows the model to dynamically utilize both semantic and detail features in layers of different depths.

The proposed STI layer allows for sufficient interaction between different frame features in GIFs, while also maintaining details between images, thus alleviating problems in modeling discrete data that traditional video generation methods may encounter.
Subsequent experiments will replace the temporal layers in VideoLDM with STI layers, so as to demonstrate the effectiveness of the proposed method in modeling discrete data.

\subsection{Experimental Setups}

\noindent
\textbf{Datasets.}
In the experiment, the cartoon and real GIFs in VSD2M are used separately as training sets, with VSD-C and VSD-R leveraged as the respective test sets.

\noindent
\textbf{Models.}
During training and inference, we fix the frame number to 8 and the output size to $256\times256$.
For stickers with less than 8 frames, we use a flip strategy to supplement frames.
For the diffusion-based methods, we use DDIM \cite{ddim} for accelerated sampling, and the number of sampling steps is 25.
The base text-image generation model is initialized from a pretrained Stable Diffusion v2.1 \footnote{huggingface.co/stabilityai/stable-diffusion-2-1}, and we train the 3D convolution layer so as to learn temporal information.
During training, the caption format has been fixed to ``\texttt{\textcolor{orange}{[Domain]} \textbackslash t \textcolor{orange}{[Description]}}''.

\noindent
\textbf{Metrics.}
We follow VideoLDM \cite{blattmann2023align} to utilize FVD \cite{unterthiner2018towards} and CLIP similarity \cite{wu2021godiva} to evaluate our models from distribution consistency and semantic correlation, where the candidates for CLIP similarity are all texts in the test set.
In addition, we also use the video quality assessment (VQA) \cite{wu2022fast} to evaluate video quality from contextual relations.



\subsection{Performance Difference on Different Datasets}

        


\subsection{Method Evaluation}

\begin{table*}[!t]
 \small
 \centering
\resizebox{0.8\linewidth}{!}{
\begin{tabular}{c|c|cccccc}
\hline
Dataset                & Metric & VideoGPT & Make-A-Video & VideoLDM & VideoFactory  & I2VGen-XL & Ours \\ \hline
 & VQA $\uparrow$    &    0.189    &   0.475           &     0.436     &        0.480 & \underline{0.481} & \textbf{0.487}     \\

VSD-R & FVD $\downarrow$    &   2645.02     &          2661.99  &    2629.48       &       2638.97   & \underline{2623.47}  & \textbf{2613.41} \\
                       & CLIP $\uparrow$  & -       &   0.213           &   0.334       &     0.223       & \textbf{0.349} & \underline{0.345}  \\ \hline
 & VQA $\uparrow$   &    0.287    &     0.378          &     0.457    &        \underline{0.495}  & 0.493   & \textbf{0.511} \\
VSD-C & FVD $\downarrow$   &    5887.49    &            5846.75  &    5822.87      &         5611.73   & \underline{5570.24} & \textbf{5513.64} \\
                       & CLIP $\uparrow$  & -        &            0.159   &       0.1690  & 0.139        & \textbf{0.174} & \underline{0.172}   \\ \hline
\multicolumn{2}{c|}{Infer Times (s)}  &  193.34      &          \textbf{11.34}      &       11.64  &       11.87 & 13.12 & 11.71 \\ 
\multicolumn{2}{c|}{Param. (B)}  &   \textbf{0.102}     &    1.835       &        1.835  &       1.847    & 2.489 & 1.856 \\
\hline
\end{tabular}
}
 \caption{Quantitative comparison of different methods by split real and cartoon for training. The inference time is evaluated on inputs of shape $1 \times 8 \times 256 \times 256$ with a single V100.
 }
\label{split_train_results}
\end{table*}

\noindent
\textbf{Automatic Metrics.}
Table \ref{split_train_results} shows the test results when real and cartoon data in VSD2M are separated for training.
\textbf{Bold} and \underline{underline} indicate the best and the second-best, respectively.
Since VideoGPT can only be controlled with tags, we only calculate VQA and FVD values for it.
When different domains are separated for training, our method achieves the best in VQA and FVD metrics, which is attributed to the detail preservation and semantic enrichment of the STI layer.
The semantic interactions of STI enable the results with more variable patterns to improve the FVD metric, which measures the distribution of source data and generated data.
While the detail preservation branch achieves an improvement in the VQA, which measures the quality of the visual information.
In terms of the CLIP metric that measures the similarity of images and texts, our method is in a suboptimal position.
The sub-optimal CLIP metric may be due to the large differences between the generated frames, thus leading to poor cross-modal similarity.
For other methods, I2VGen-XL performs better in terms of visual quality.
VideoGPT performs poorly on both VQA and FVD metrics and is still far from the diffusion-based approaches.
In addition, we also provide the joint training results of the cartoon and real part in the Appendix \textcolor{red}{A.2}.

Fig. \ref{methods_training_change} shows the indicator changes of different models as the number of training steps increases.
In the later stages of training, our model performs better than other models on VQA and FVD indicators, which verifies that our model has stronger fitting ability.

\begin{figure}[!h]
\centering
\includegraphics [width=3.0in]{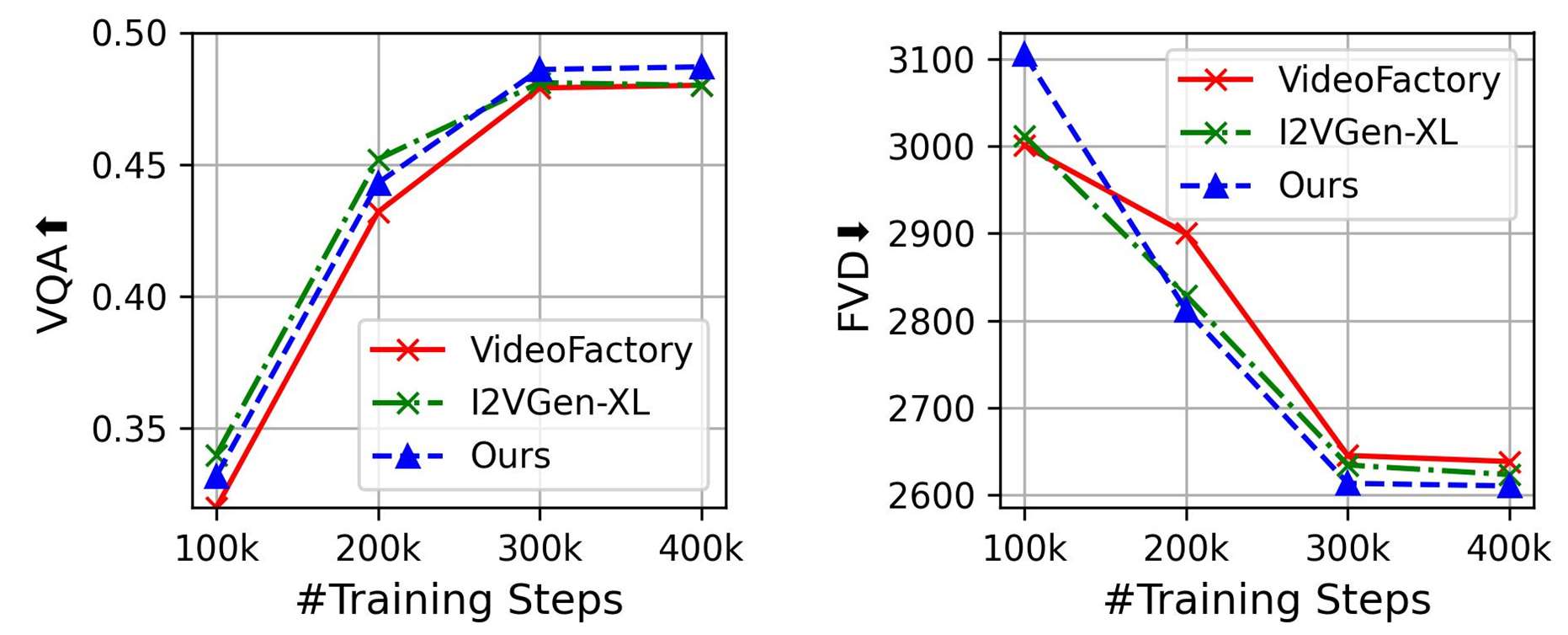}
\captionsetup{font={small},skip=4pt, justification=raggedright}
\caption{
The indicator changes of different models as the number of training steps increases.
}
\label{methods_training_change}
\end{figure}

The inference time in Table \ref{split_train_results} refers to the duration between one sample process.
The experiments are performed on a single V100 graphics card with no load to sample 8-frame GIFs in 256 size, and each set of experiments is conducted hundreds of times to obtain average results.
Although the transformer-based method has a smaller parameter size, it has no advantage over the diffusion-based method in terms of inference time simply due to serial prediction.
For the diffusion-based method, VideoFactory is at a disadvantage in both two metrics, and the reason is that the introduction of temporal and spatial attention in each UNET block.
Although our model introduces the STI layer compared to VideoLDM, the number of parameters and time consumption are basically the same as the diffusion-based model.

\noindent
\textbf{Qualitative Comparison.}
layerll
The above method can obtain results with clearer front and background, with less subject distortion when coping with the ASG task.
In addition, in experiments, we found that compared to other video generation models based on diffusion methods, our method can produce more discrete GIF results with an stronger sense of motion.


\begin{figure}[!h]
\centering
\includegraphics [width=3.2in]{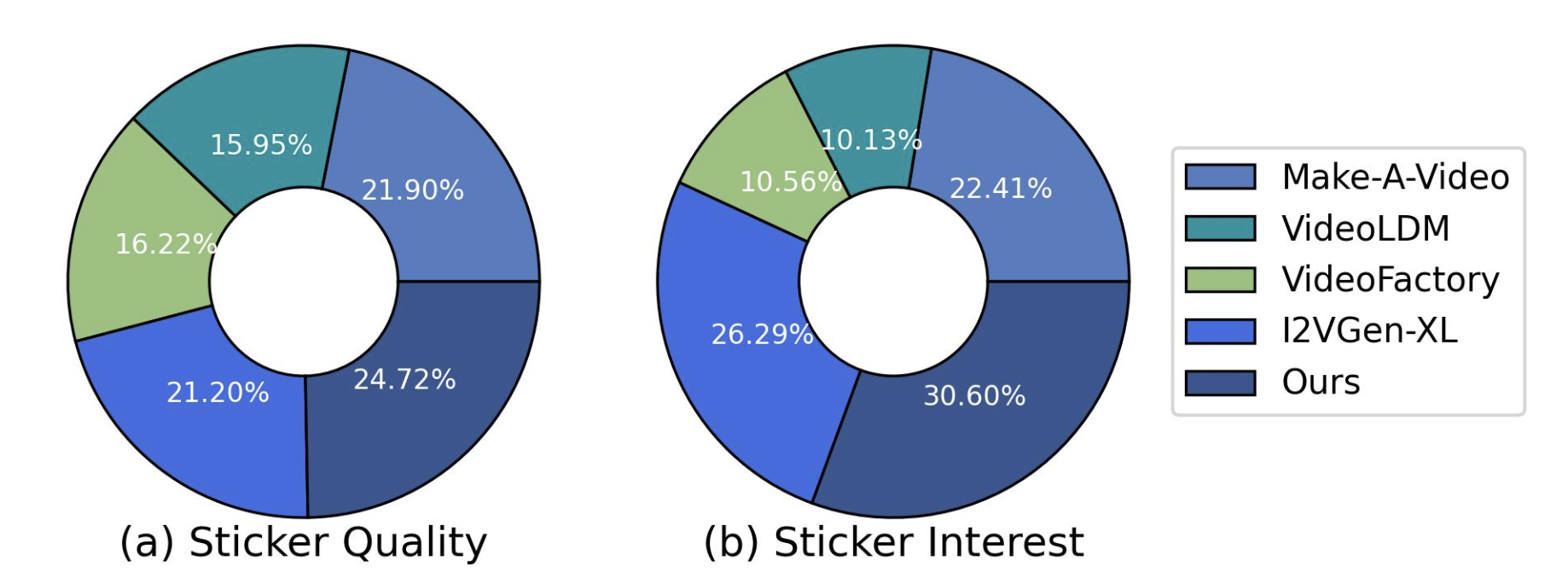}
\captionsetup{font={small},skip=4pt, justification=raggedright}
\caption{Manual comparison of sticker generated by different methods on (a) quality and (b) interest.}
\label{pie}
\end{figure}

\noindent
\textbf{User Preference.}
In addition to automatic metrics, we also introduce manual evaluation to overcome the limitation of existing metrics.
Since VideoGPT cannot be controlled by semantics, we only conduct relative comparisons with the other three methods.
We request nine labelers to select the best generated stickers in terms of quality and interest, and the related results is shown in Fig.\ref{pie}(a) and Fig.\ref{pie}(b), respectively.
In subjective evaluation, our method achieved the best visual quality, followed by Make-A-Video, which shows that our model has better diversity when generating animated stickers.
In terms of user interest, our model is significantly superior to other models, possibly because the results generated by our method have a higher degree of discretization with richer semantics.
Additionally, we provide a more detailed subjective evaluation result in Appendix \textcolor{red}{A.2}.

\begin{table}[!h]
\centering
\resizebox{\linewidth}{!}{
\begin{tabular}{c|cc|cc}
\hline
\multirow{2}{*}{Training Sets} &  \multicolumn{2}{c}{VSD-R} & \multicolumn{2}{c}{VSD-C}   \\
& VQA $\uparrow$ & FVD $\downarrow$ & VQA $\uparrow$ & FVD $\downarrow$ \\
\hline
TGIF \cite{li2016tgif}  & 0.394    &           3310.25                                            &    -                        & - \\
SER30K \cite{liu2022ser30k}  &      -                      &                     -    &    0.372  &             5813.47                             \\
VSD2M    &                                   \textbf{0.487} & \textbf{2613.41}   &    \textbf{0.511}           &                      \textbf{5513.64}        \\ \hline
\end{tabular}
}
 \caption{Performance comparison of training models under different training sets.
}
\label{tgif}
\end{table}

\noindent
\textbf{Compare to Other Datasets.} 
To verify the strengths of VSD2M compared with other datasets, we use our method to train or fine-tune on different datasets and test on VSD-R and VSD-C testsets.
As shown in Table \ref{tgif}, the VQA and FVD indicators of the model trained on VSD2M have been greatly improved compared with others such as TGIF\cite{li2016tgif} and SER30K\cite{liu2022ser30k}.
This experiment verifies that VSD2M has the ability to generate the well stickers compared with other datasets.




Overall, the current diffusion-based methods demonstrate superior performance on ASG tasks, yet they still fall short in metrics such as image quality and semantic relevance, which are discussed in Appendix \textcolor{red}{A.3}. 
If researchers can further design models based on the inherent properties of GIFs, it may be able to greatly promote the development of the ASG field.

\section{Conclusion}
\label{Conclusion}

This paper firstly proposes a million-level vision-language sticker dataset to facilitate the task of animated sticker generation.
VSD2M covers both static and animated stickers, which is currently the largest dataset in sticker field.
Furthermore, we come up with a spatial temporal interaction layer and establish a set of solid baselines including transformer-based and diffusion-based approaches to provide exhaustive benchmarks for ASG tasks.
As far as we know, this is a well-established work for animated sticker generation, and we hope that this work can serve as inspiration and guidance for researchers in the field of intelligent creation.

\bibliography{aaai25}

\clearpage

\section{\LARGE{\textcolor{black}{Appendix}}}
\section{A.1 Selection of Comparison Methods}

\begin{table}[!htbp]
\resizebox{\linewidth}{!}{
\begin{tabular}{cccc}
\hline
Method              & Publication & Category  & Institution         \\ \hline
VideoGPT \cite{yan2021videogpt}         & arXiv'21   & Transformer-based & Berkeley \\
Make-A-Video \cite{singer2022make}        & ICLR'23   & Diffusion-based & Meta AI  \\ 
VideoLDM \cite{blattmann2023align}           & CVPR'23   & Diffusion-based  & NVIDIA  \\
VideoFactory \cite{wang2023videofactory}        & arXiv'23   & Diffusion-based & Microsoft \\
I2VGen-XL \cite{zhang2023i2vgen} & arXiv'23 & Diffusion-based & Alibaba \\
\hline
\end{tabular}}
\caption{Selected state-of-the-art video generation methods under different types.}
\label{methods_comp}
\end{table}

\begin{table*}[!t]
 \centering
\begin{tabular}{c|c|cccccc}
\hline
Dataset                & Metric & VideoGPT & Make-A-Video & VideoLDM & VideoFactory & I2VGen-XL & Ours \\ \hline
 & VQA $\uparrow$    &   0.226     &  0.491            &    0.454      &       \underline{ 0.494} & 0.483 &  \textbf{0.503}    \\

VSD-R & FVD $\downarrow$    &       2780.44 &            2636.68  &    2589.86      &       2605.70 & \underline{2573.82} & \textbf{2550.11}     \\
                       & CLIP $\uparrow$  &  -      &            0.204  &   0.311  &    0.266   & \textbf{0.319} &    \underline{0.314}   \\ \hline
 & VQA $\uparrow$   &   0.233     &          0.447    &      0.457    &         0.535    & \underline{0.552} & \textbf{0.557} \\
VSD-C & FVD $\downarrow$   &   5731.82     &         5824.31     &    5822.32      &      5716.40 & \underline{5621.12} &  \textbf{5503.57}    \\
                       & CLIP $\uparrow$  &  -      &  0.185            &   0.191       &   0.126   & \textbf{0.201} &  \underline{0.193}     \\ \hline
\end{tabular}
 \caption{Quantitative comparison of different methods by joint real and cartoon for training.
}
\label{joint_train_results}
\end{table*}

\begin{figure}[!h]
\centering
\includegraphics [width=3.2in]{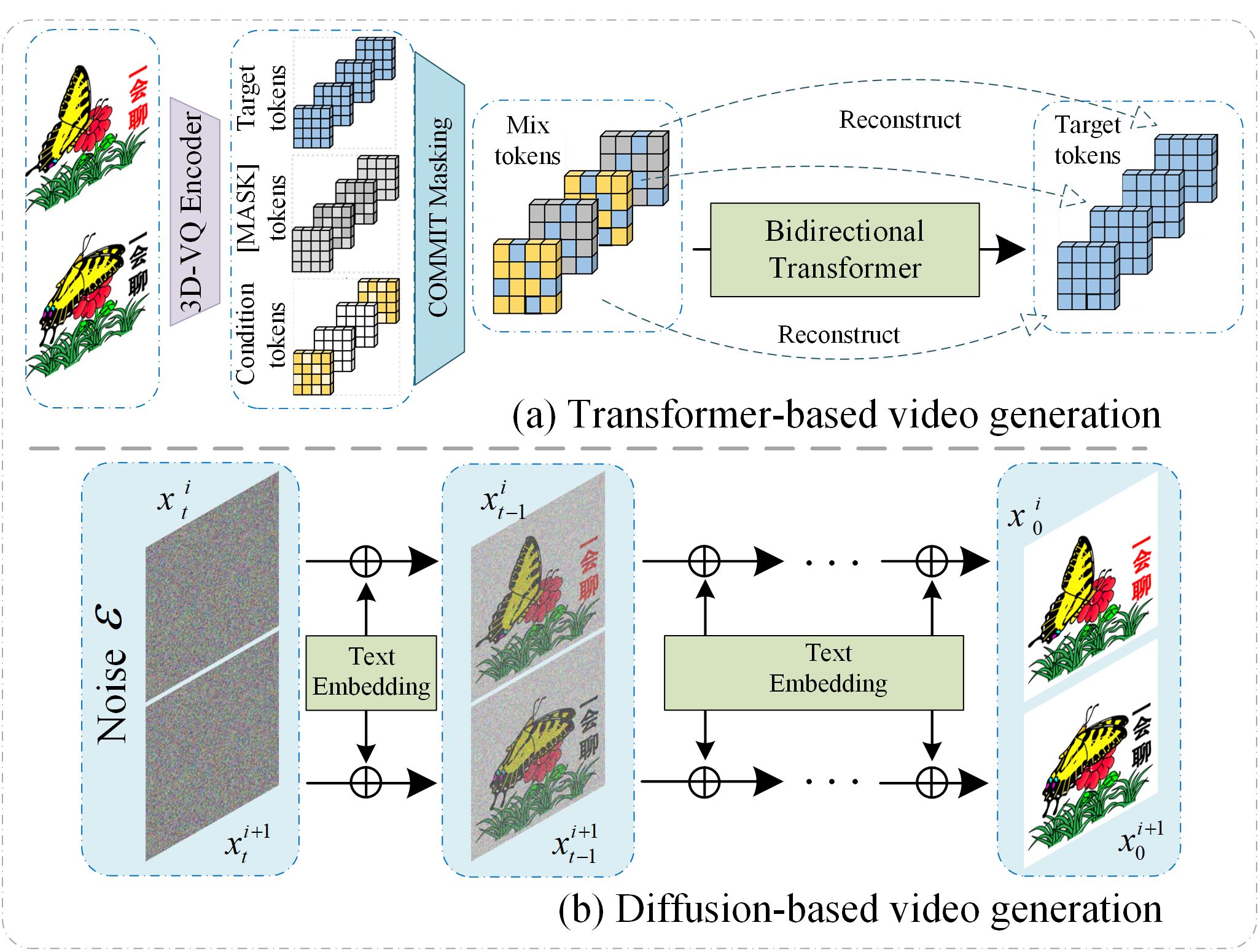}
\captionsetup{font={small},skip=4pt, justification=raggedright}
\caption{
Framework comparison of transformer-based and diffusion-based video generation.
The transformer-based method adds multiple masks to 3D tokens so as to perceive the video distribution, while the diffusion-based method leverages score matching to achieve distribution transfer from Gaussian noise to continuous frames.
}
\label{methods_divide}
\end{figure}

As shown in Table \ref{methods_comp}, 
We select four state-of-the-art models from transformer-based and diffusion-based video generation methods, where the difference between the two architectures is shown in Fig. \ref{methods_divide}.
We aim to model the ASG task using different types of research methods, so as to analyze the feasibility of each method.

\section{A.2 Additional Evaluation Results}

\begin{figure}[!t]
\centering
\includegraphics [width=3.2in]{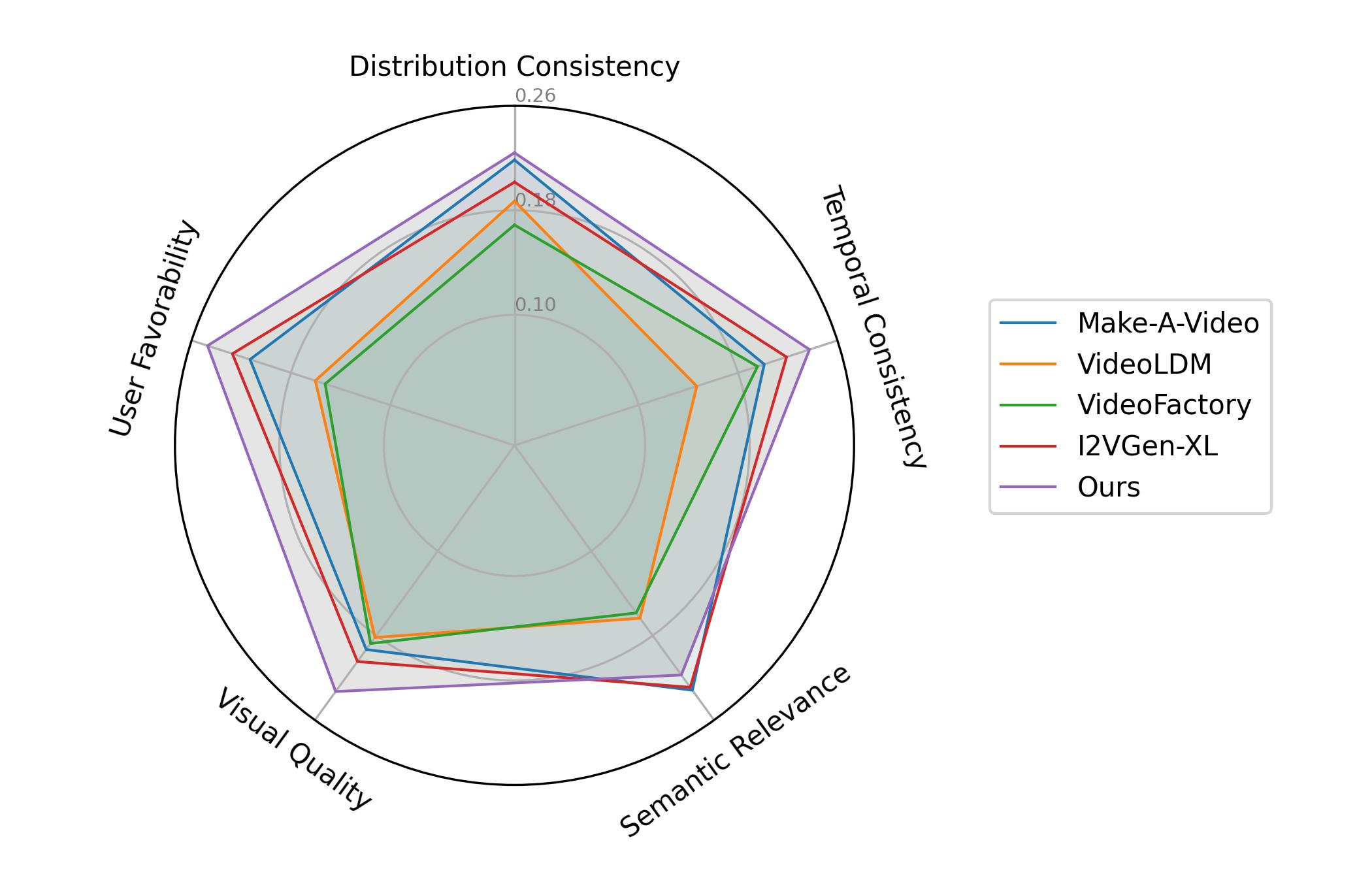}
\captionsetup{font={small},skip=4pt, justification=raggedright}
\caption{
More detailed subjective evaluation results, which are evaluated from the perspectives of temporal consistency, semantic relevance, visual quality, distribution consistency, and user favorability.
}
\label{radar}
\end{figure}

\textbf{Comprehensive Subjective Evaluation.}
In order to evaluate the generated results more comprehensively, we request ten people to rank the generated results from the following aspects:
\textbf{(a) Distribution consistency.} Action and scene similarity between the generated sticker and the reference sticker.
\textbf{(b) Temporal consistency.} Inter-frame consistency of generated stickers.
\textbf{(c) Semantic relevance.} The correlation degree between the generated sticker and the caption provided in the test set.
\textbf{(d) Visual quality.} Whether the generated sticker has a mutilated subject or a collapsed image.
\textbf{(e) User favorability.}
User preference on generated sticker.
We count the number of optimal samples for each method and summarize the results of the above five aspects in Fig. \ref{radar}.
Our method is in the leading position in temporal consistency, visual quality, and user favorability, which benefits from the STI layer’s modeling characteristics of discrete data.
Secondly, our method is slightly inferior to I2VGen-XL in terms of semantic relevance, which is similar to the CLIP indicator in the quantitative results.
Compared with VideoLDM, VideoFactory is superior in temporal consistency, which verifies spatial-temporal attention mechanism \cite{wang2023videofactory} in sticker generation.
While compared with Make-A-Video, VideoLDM has a small improvement in distribution consistency, and the difference between the other three indicators is not significant.

\noindent
\textbf{More Automatic Results.}
Table \ref{joint_train_results} shows the quantitative comparison of different methods by joint real and cartoon for training.
Our method performs well in VQA and FVD indicator and is superior to I2VGen-XL in image-text similarity.


\section{A.3 Challenges and Future Works}
\label{challenges}

As a novel task that urgently needs to be explored, ASG task plays an important role in user interaction and chat communities.
However, compared with video generation task in general fields, ASG task present some unique challenges:
\begin{itemize}
    \item Due to the need for manual creation and copyright issues, the collection of emoticons is greatly limited and can only be obtained from a small number of websites.
    Few sticker samples are difficult to train a robust generative model, for this reason the fine-tuning methods such as PEFT \cite{peft} need to be introduced to reduce the
    demand for samples.
    \item 
    Sticker usually covers a large number of scenes ($i.e.$ cartoon and real scenes), while for a specific scene, they contain action, characters, $etc.$ that are often interesting and rare for generic scenes.
    On the other hand, stickers always contain a large amount of optical characters to highlight the theme, and the distribution of these texts is messy and difficult to model.
    In this scenario, the generative model may need to adapt to scaling law by employing more parameters to handle such a distribution.
    \item
    For animated stickers, the diversity and abstraction of content makes it difficult to divide the actions at a fine-grained level, which makes the model lacks perception of motion during learning.
   In addition, stickers are difficult to describe. For example, a dog in one sticker is greatly different from another, and the text cannot be fully used to better distinguish.
   How to control the fine-grained subject and motion in the sticker is one of the urgent issues that need to be studied.
\end{itemize}

\noindent
Under the above challenges, there are also some corresponding opportunities in ASG field:
\begin{itemize}
    \item Collecting larger-scale data and using self-supervised or supervised methods to obtain pre-trained models for ASG task is an important milestone.
    By learning the common features in stickers, the pre-trained model can quickly iterate on downstream tasks, thus greatly promoting the development of this field.
    \item The cartoon stickers usually consist of simple lines and color blocks, with much less texture than the nature scene.
    While for lines and blocks, the two distributions are extremely different in the frequency domain, and it may be necessary to disassemble and reconstruct them separately.
    \item The creation of artificial stickers is often based on the process of sketching-coloring, and how to model ASG task based on this process is also a promising direction.
    Whether the sticker generation can be broken down into sketching and coloring, thereby reducing the modeling difficulty and improving the sample quality, is an interesting approach that needs to be explored.
    \item
    Similar to natural scenes, how to achieve detailed control in the generation process is also an indispensable part in intelligent creation.
    Characterization of subjects and modeling of action will inevitably become one of the bottlenecks in generating high-quality stickers.
\end{itemize}

\end{document}